\documentclass[fleqn,a4paper,10pt]{article}
\usepackage{amsmath}
\usepackage{enumerate}
\usepackage{enumitem}  
\usepackage[hmargin=2.5cm,vmargin=3.5cm]{geometry}
\usepackage{xcolor}
\usepackage[colorlinks = true,
            linkcolor = blue,
            urlcolor  = blue,
            citecolor = blue,
            anchorcolor = blue]{hyperref}
\usepackage{multirow}
\usepackage{graphicx}
\usepackage{dcolumn}
\usepackage{bm}

\begin{document}

\title{Novel doping alternatives for single-layer transition metal dichalcogenides}
\author{Nicolas Onofrio$^1$\footnote{Corresponding author: nicolas.onofrio@polyu.edu.hk}, David Guzman$^2$ and Alejandro Strachan$^2$}
\date{}

\maketitle

\noindent
$^1$ Department of Applied Physics, The Hong Kong Polytechnic University, Hong Kong SAR \\
$^2$ School of Materials Engineering and Birck Nanotechnology Center Purdue University, West Lafayette, IN 47906 USA \\

\begin{abstract}
Successful doping of single-layer transition metal dichalcogenides (TMDs) remains a formidable barrier to their incorporation 
into a range of technologies. We use density functional theory to study doping of molybdenum and tungsten dichalcogenides 
with a large fraction of the periodic table. An automated analysis of the energetics, atomic and electronic structure of thousands of 
calculations results in insightful trends across the periodic table and points out promising dopants to be pursued experimentally. 
Beyond previously studied cases, our predictions suggest promising substitutional dopants that result in p-type transport and reveal 
interesting physics behind the substitution of the metal site. Doping with early transition metals (TMs) leads to tensile strain and a 
significant reduction in the bandgap. The bandgap increases and strain is reduced as the d-states are filled into the mid TMs; these 
trends reverse are we move into the late TMs. Additionally, the Fermi energy increases monotonously as the d-shell is filled from the 
early to mid TMs and we observe few to no gap states indicating the possibility of both p- (early TMs) and n- (mid TMs) type doping. 
Quite surprisingly,  the simulations indicate the possibility of interstitial doping of TMDs; the energetics reveal that a significant 
number of dopants, increasing in number from molybdenum disulfide to diselenide and to ditelluride, favor the interstitial sites over
adsorbed ones. Furthermore, calculations of the activation energy associated with capturing the dopants into the interstitial site indicate 
that the process is kinetically possible. This suggets that interstitial impurities in TMDs are more common than thought to date and 
we propose a series of potential interstitial dopants for TMDs relevant for application in nanoelectronics based on a detailed analysis 
of the predicted electronic structures.
\end{abstract}

\newline

\section{Introduction}

Transition metal dichalcogenides (TMDs) are among the most promising candidates to replace silicon in next-generation, ultrascaled electronic
devices. Attractive features of TMDs include  a wide range of chemical compositions, phases, as well as electronic properties that depend 
not just of chemistry but also on the number of layers~\cite{wang2012electronics,chhowalla2013chemistry}.
Breakthroughs in synthesis have enabled a variety of  nanodevices~\cite{jariwala2014emerging} including field-effect transistors~\cite{radisavljevic2011single},  
memories~\cite{bessonov2015layered}, and sensors~\cite{li2012fabrication}; even an entire electronic circuit has been proposed based on TMDs and graphene~\cite{yu2014graphene}. However, despite the significant promise of TMDs for electronic applications, surpassing silicon-based devices in 
performance and reliability is challenging and will require progress on several fronts; doping is key among them.

As in bulk semiconductors, controlled doping levels and charge carrier concentration in single-layer TMDs is critical to enable digital electronics.
Two major doping strategies have been pursued theoretically and experimentally: {\it chemical} and {\it substitutional} doping. 
Chemical doping consists of the adsorption of atoms or molecules on the surface of the TMD leading to the alteration 
of its electronic structure as a consequence of surface charge transfer. Javey and coworkers have pioneered this approach and demonstrated 
p-type conduction in NO$_2$ doped WSe$_2$ field effect transistors. The chemisorption of NO$_2$ acts as surface electron ``pumps'' and not only 
affects carrier density but also dramatically lowers the contact resistance with various metals~\cite{fang2012high}.
The second approach, substitutional doping of TM and chalcogen atoms in TMDs, has also been recently demonstrated 
experimentally. For example, p-type conduction has been measured in MoS$_2$ by substituting Mo with Nb~\cite{suh2014doping} or S with 
N~\cite{azcatl2016covalent} while strong n-type conduction has been reported when S is substituted by Cl~\cite{yang2014chloride}. 
These doping effects are consistent with simple models based on the change in the total number of electrons in the compound by selecting a dopant 
having one more or one less electron in its valence shell than the substituted atom. 
This has been confirmed from first principles calculations showing the Fermi level shifting of chalcogen substituted TMDs with halogen and group V 
elements~\cite{yue2013functionalization} as well as TM substitution with Nb~\cite{dolui2013possible}. Substitutional doping can induce 
strain to the TMDs~\cite{azcatl2016covalent} and, recent calculations indicate that phase transitions can occur upon doping~\cite{raffone2016mos2}. 
An additional challenge is that defects and the location of conduction and valance bands results in intrinsic doping of TMDs, while this is difficult to compensate for
~\cite{suh2014doping}, recent research suggests the possibility of healing these defects via oxygen passivation ~\cite{lu2015atomic}; 
resulting in substitutional doping. 

Substitutional doping has been studied via electronic structure calculations; yet previous studies are limited in the number of substituents studied ~\cite{dolui2013possible,mishra2013long,cheng2013prediction,lu2014electronic,hu2015electronic} and there is a need to identify doping trends across 
the periodic table and across the chemistry of TMDs. 
We believe such knowledge will contribute to the development of more efficient doping strategies, especially for p-type. 
Perhaps surprisingly, the possibility of interstitial doping in single-layer TMDs has not been explored either experimentally or theoretically;
to our knowledge, only intrinsic interstitial defects have been reported in single-layer TMDs~\cite{haldar2015systematic}. 
More generally, it is not clear whether 2D materials in general, from graphene~\cite{geim2007rise} and boron nitride~\cite{chopra1995boron} to
emerging {\it puckered} 2D materials like the Xenes family (silicene~\cite{vogt2012silicene}, gemanene~\cite{ni2011tunable}, 
and stanene~\cite{zhu2015epitaxial}), phosphorenes~\cite{liu2014phosphorene} can support interstitial doping.
Previous studies show interstitial sites energetically favorable in Xenes~\cite{sahin2013adsorption,manjanath2014mechanical,xing2017tunable} 
however this has not been predicted in TMDs.

\begin{figure}[!ht]
\centering
\includegraphics[width=0.65\textwidth]{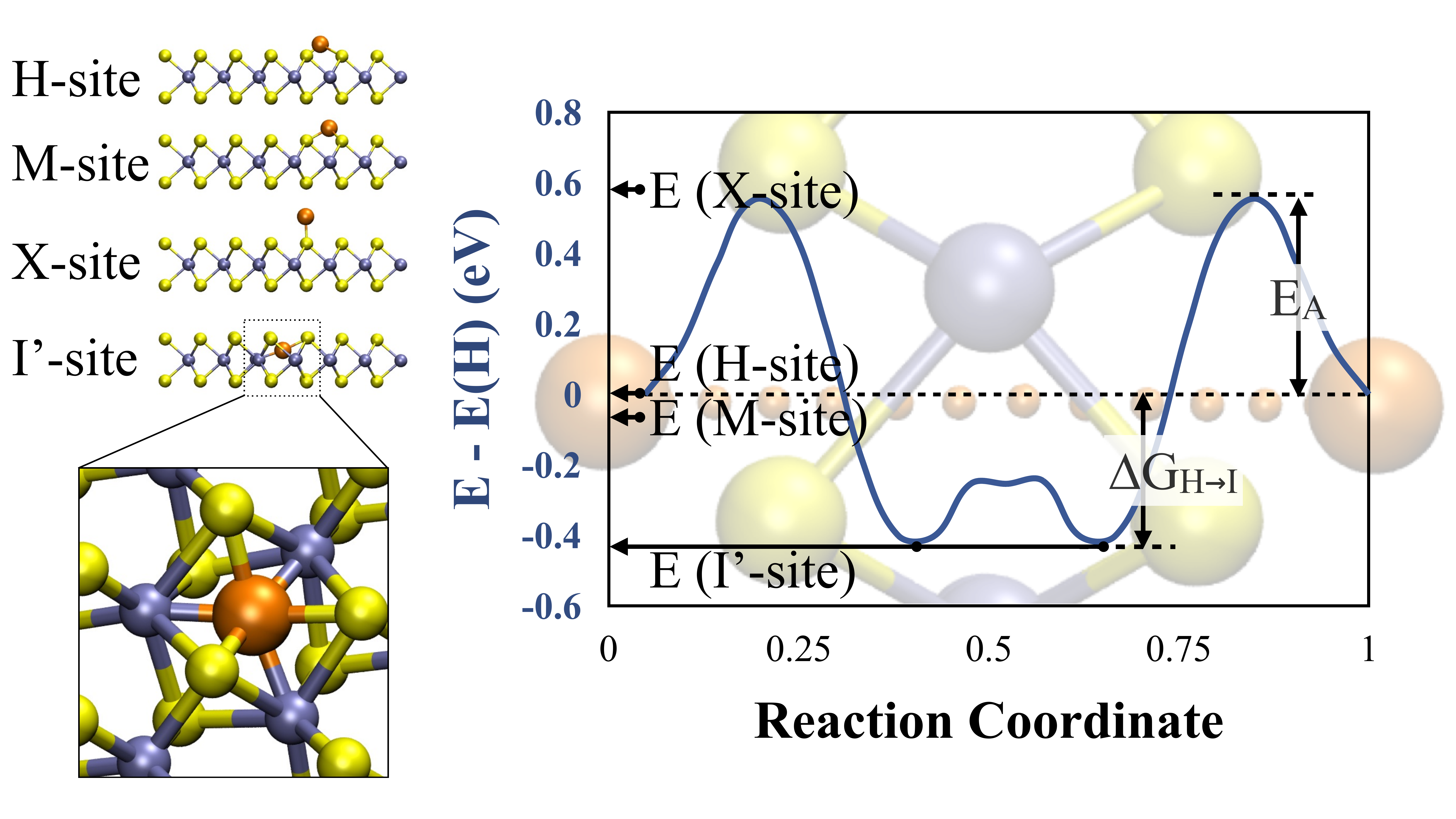}
\caption{Side-view of a Cu atom adsorbed at various locations (H, M and X) on the surface of a free standing trigonal prismatic TMD and at interstitial (I') site (left).
Potential energy surface of Cu crossing a single-layer of MoTe$_2$ (right).
We define the activation energy $E_A$ and the free energy $\Delta G_{H\rightarrow I}$ of the doping reaction.}
\label{fig:fig1}
\end{figure}

In this paper we use high-throughput density functional theory (DFT) to characterize substitutional and interstitial doping of Mo and W dichalcogenides with
elements encompassing a large fraction of the periodic table. Simulation details are presented in Section~\ref{sec:sim}. We characterize the energetics, atomic 
structure and electronic properties as a function of the dopant. In both substitutional and interstitial doping, interesting trends in energetics and electronic 
properties emerge across the periodic table.
In addition to known substitutional dopants, we show in Section~\ref{sec:sub} that B, Ge and Sn substitution of the chalcogen atom are 
promising for p-doping, so is the substitution of the metal site with Ti, Zr, Si, Ge, Sn and B. 
N-type doping is more challenging and besides the previously studied dopants we find that V and Nb substitution of the chalcogen atom and Tc doping of 
Mo show significant promise. 
Figure \ref{fig:fig1} illustrates the idea of interstitial doping of single layer TMDs; it shows the molecular structure a Cu atom at various adsorption sites 
on a single-layer of MoTe$_2$ (trigonal prismatic structure) and the configuration corresponding to an interstitial site. 
The right panel of Figure \ref{fig:fig1} shows the minimum energy path as a Cu atom is marched across the freestanding TMD. 
We can clearly see that the lowest energy corresponds to the Cu atom at an interstitial site (I'-site), in a distorted octahedral configuration formed with 3 Mo and 3 Te neighbors.
Interestingly we find that a number of possible dopants prefer energetically the interstitial position over the adsorbed sites, the number increasing
as we move from sulfides to selenides and tellurides. Section~\ref{sec:int} describes the trends in energetics and electronic properties of interstitially doped TMDs.

\section{Simulation details}
\label{sec:sim}

\subsection{DFT details} 

High-throughput  DFT calculations were performed with the SeqQuest package developed at Sandia National Laboratories~\cite{seqquest} within the generalized gradient 
approximation proposed by Perdew, Burke, and Ernzerhof (PBE)~\cite{perdew1996generalized}. 
We included Grimme's D2 correction to compute the energy of adsorbed and interstitial positions for a better treatment of London dispersion. 
Supercells consist of monolayer TMDs taken in their ground state trigonal prismatic (H-phase) replicated 4$\times$4 times along the in-plane lattice directions with
more than 30 \AA~of vacuum space between layers.
Doping with a single atom corresponds to a concentration of approximatively 7$\times$10$^{13}$ cm$^{-2}$, in agreement with common doping concentration 
for field effect transistor application~\cite{suh2014doping,azcatl2016covalent}.
All structures were fully relaxed including lattice parameters and ionic positions until energies, forces and pressure reach values of 0.004 eV, 0.08 eV/\AA~and 
0.2 N/m$^2$, respectively. 
Note that this is in contrast with common practice in bulk materials where the lattice parameters are usually not relaxed due to lower dopant concentration~\cite{freysoldt2014first}.
A 2$\times$2$\times$1 k-grid was used for integrals in reciprocal space in all simulations.  
The robust Broyden atomic and cell relaxation algorithm~\cite{broyden1965class} implemented in SeqQuest is key to our methodology, as we found high success 
rate in the structural relaxations with the simulation achieving the convergence criteria within a relatively small number of steps.
Moreover, SeqQuest automatically and exactly eliminates dipole interactions between periodic slab images using the local moment counter charge 
method~\cite{schultz1999local}.
In order to include potential magnetic effects, each doped TMD has been relaxed at various spin states up to 3 (corresponding to spin polarizations 1 to 6).
To assess the accuracy of the level of theory, we re-computed selected doped TMDs using the state-of-the-art Heyd, Scuseria and Ernzerhof (HSE) exchange 
correlation potential~\cite{heyd2003hybrid}, as implemented in VASP~\cite{kresse1996efficiency}. HSE mixes exact exchange and has been 
shown to provide an accurate description of atomic and electronic structures of some TMDs~\cite{peelaers2014first,ramasubramaniam2012large}; unfortunately, the
computational intensity of the method (typically 30 fold increase in compute time) preclude its use to relax all the configurations studied in this papers.
Finally, we did not include spin-orbit coupling to the present study since we found negligible effect on the structural relaxation and energetics of various doped TMDs 
(see Section S3).

\subsection{Doping procedure} 

The elements tested include the first five rows of the periodic table, corresponding to 54 elements.
Substitutional doping was achieved at the metal and chalcogene sites in Mo and W disulfides and diselenides.
We did not include ditellurides to the substitutional doping analysis because of the large fraction of the doped TMDs leading to ground state 
distorted octahedral structure, as it has been observed previously~\cite{raffone2016mos2}.
In order to evaluate potential interstitial dopants, we compared the energy of elements adsorbed on the surface of the monolayer with that at interstitial sites.
Possible adsorption sites, on the surface of trigonal prismatic TMD, are located on top of a transition metal (M-site), on top of a chalcogen atom (X-site) 
and on top of a hollow site (H-site), as detailed Figure S7.
The equilibrium distance between the dopant and the TMD varies with the nature of the TMD and the element adsorbed. 
We found the best initial locations to be at approximately 0.5 \AA~and 1.75 \AA~away from the chalcogen atoms at H (as well as M) and X sites, respectively.
We identified two interstitial sites, both with in-plane positions identical to those of the hollow site and located in the symmetry plane of the 
TMD (I-site) or slightly shifted from the symmetry plane (denoted I'-site and seen for Cu in Figure \ref{fig:fig1}).
For validation purpose, we analyzed the final relaxed atomic structures to confirm whether the dopant remained in its doping site; 
we found that alkali metals and noble gas do not interact strongly with the host TMD therefore, we will not consider them as substituants in the following analysis.
Figure S1 and S8 summarize the final (after relaxation) positions of the dopants with respect to the TMD symmetry plane upon substitutional and
interstitial/chemical doping, respectively.

\subsection{Simulation analysis} 

The formation energy associated with metal or chalcogen substitution by dopant $D$ is defined as:

\begin{equation}
\label{eq:eq1}
E_{D}^{M/X} = \left[ E(nMX_2+D^{M/X}) + \mu_{M/X} \right] -   \left[ nE_{MX_2} + \mu_{D} \right]
\end{equation}

where $E(nMX_2+D^{M/X})$ is the energy of a periodic system containing $n$ formula units including dopant D taking the 
site of the metal or chalcogen (denoted by superscript M/X), $E_{MX_2}$ is the energy of a perfect crystal TMD per formula units 
and $\mu$ denotes the chemical potential of various species, either M, X or D. 
In the case of interstitial doping the formation energy is defined as:

\begin{equation}
\label{eq:eq2}
E_{D}^{I} = \left[ E(nMX_2+D^{I})\right] -  \left[ nE_{MX_2} + \mu_{D} \right]
\end{equation}

where $E(nMX_2+D^{I})$ is the energy of the TMD sample with $n$ formula units including an interstitial dopant. 
We note that the formation energy for an adsorbed atom is defined following Eq.~\ref{eq:eq2} as well.
It is clear from Eqs.~\ref{eq:eq1} and~\ref{eq:eq2} that the formation energies depend on the chemical potential of the species involved,
which are determined by growth conditions. The chemical potentials for the metal and chalcogen atoms are typically considered between
two limits: metal rich and chalcogen rich. Under metal-rich conditions \{$\mu_M^{M-rich}$,$\mu_X^{M-rich}$\}, the chemical potential of the 
metal is set by its ground state crystal structure and that for the chalcogen is set  such that the TMD is in equilibrium with the metal source.
In chalcogen-rich conditions \{$\mu_M^{X-rich}$,$\mu_X^{X-rich}$\}, the chemical potential of the chalcogen is obtained from the di-atomic 
molecule  and that of the metal obtained assuming equilibrium with the TMD~\cite{dolui2013possible}. Thus, we define:

\begin{equation}
\label{eq:eq3}
  \begin{cases}
    \mu_M^{M-rich} &= \mu_M^0 \\
    \mu_X^{M-rich} &= \frac{1}{2}\left(E_{MX_2}-\mu^0_M\right)
  \end{cases}
  \qquad
  \begin{cases}
    \mu_M^{X-rich} &= E_{MX_2}-2\mu_X^0 \\
    \mu_X^{X-rich} &= \mu_X^0
  \end{cases}
\end{equation}

with $\mu_M^0$ and $\mu_X^0$ the chemical potentials of the metal and chalcogen atoms taken in their standard conditions, respectively. 
The chemical potential of the dopants $\mu_D$ were also calculated in their standard conditions.

In order to graphically showcase trends in the electronic structure across the entire periodic table, we developed an algorithm to automatically analyze
the electronic density of state (DoS) and extract conduction and valence band edges, Fermi energy and defects within the bandgap. The method starts
from a DoS obtained from Kohn-Sham eigenvalues with a gaussian smearing of 0.05 eV.
The key steps are summarized as follows:

\begin{enumerate}[label=(\roman*)]
\item Starting from the Fermi energy (E$_F$; evaluated by the DFT code from standard population analysis) we define an energy range [E$_F-\delta$,E$_F+\delta$] 
that extends well into the valence and conduction bands.
The energy parameter should satisfy: $\delta > E_{BG}$ with $E_{BG}$, the bandgap energy. For our PBE calculations, we choose the value $\delta = 2.0$ eV.
\item The DoS is scanned as a function of energy from both ends toward the Fermi energy and the band edges are defined when the density becomes less than 
a threshold energy E$_{thr}$ (discussed below). For spin-polarized calculations, we consider the averaged spin up and down densities.
\item In order to identify defects within the bandgap, the range of energies between the conduction and valence band edges is scanned for densities
larger than E$_{thr}$.
For a better detection of defect states in spin-polarized calculations, both spin channels defects are considered.
\end{enumerate}

We generated all DoS plots and performed a careful comparison with the automated band edge plot.
We provide plots of the projected DoS on our \href{https://nanohub.org/labs/run/pdos/}{website}.
The analysis offers the best resolution, including the location of band edges and defects with E$_{thr}$=0.5 eV$^{-1}$. 

\section{Substitutional doping} 
\label{sec:sub}

\subsection{Strain and energetics of doping} 

\begin{figure}[!ht]
\centering
  \includegraphics[width=0.95\textwidth]{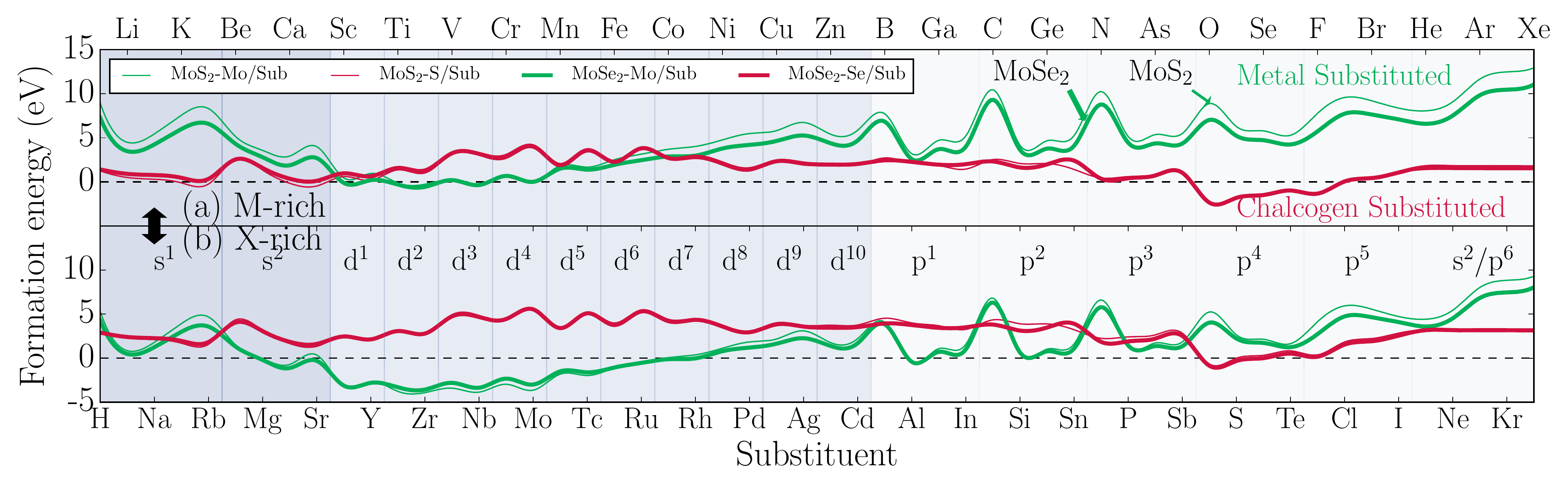}
  \caption{Formation energy of Mo-substituted (green) and chalcogen-substituted (red) MoX$_2$ (with X=S, Se) in M-rich (a: top) and X-rich (b: bottom) conditions as a function of substituent.}
  \label{fig:fig2}
\end{figure}

\begin{figure}[!ht]
\centering
  \includegraphics[width=0.95\textwidth]{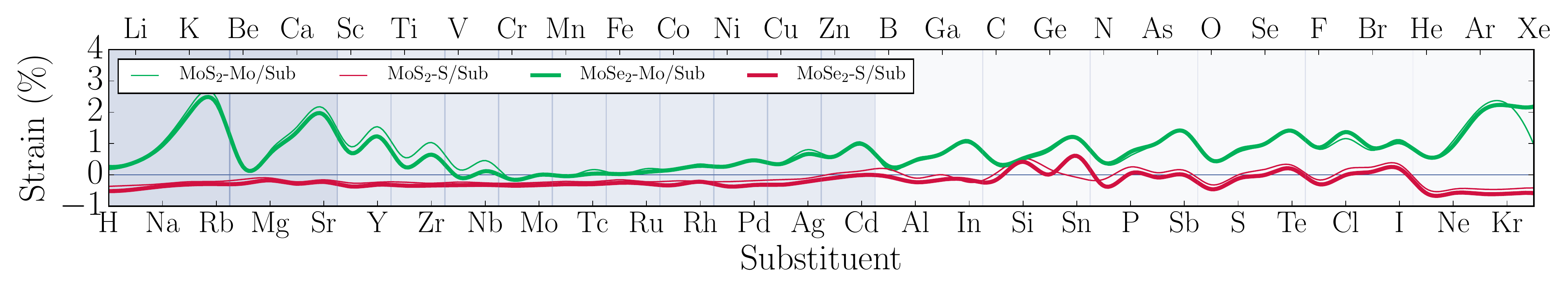}
  \caption{In-plane biaxial strain of the lattice consequence of substitutional doping of MoX$_2$ (X=S, Se) as a function of substituent.}
  \label{fig:fig3}
\end{figure}

Figures \ref{fig:fig2} and \ref{fig:fig3} show the formation energy of substituted MoS$_2$ and MoSe$_2$ calculated in metal- (M-rich) and 
chalcogen-rich (X-rich) conditions as well as the corresponding in-plane strain induced to the lattice, respectively.
The figures corresponding to WS$_2$ and WSe$_2$ are reported in the Supplementary Material, Figure S2 and S3.
Our predictions are in good agreement with previous work, as detailed Section S1.3.

Overall, M-rich conditions lead to a higher chemical potential for the chalcogen and, consequently, favor substitutional doping of the chalcogen site. 
Conversely, X-rich conditions favor TM substitutions.
Substitutional doping of Mo (or W) with TMs is predicted to be energetically feasible. 
The formation energies are relatively constant for early TMs and negative in X-rich conditions and increase from mid to late TMs. 
The associated strain is tensile for early TMs and decreases as we move into the mid TMs; the strain increases again as the d-shell is completed. 
The induced strain by the TMs and formation energy both follow the non-monotonous trend of atomic radius.
In addition to TMs, the predicted energetics show that Mo (and W) could be substituted by other metals and semiconductors except for first row 
elements (B $\rightarrow$ F). 
We attribute this effect to their localized 2p electrons being unable to make chemical bonds with the host TMD. 
Moreover, we found no correlation between Mulliken charges on the substituent and the corresponding formation energies, ruling out significant ionic
interactions.

Substitutional doping of the chalcogen atom with other chalcogens as well as group V elements and halogens is predicted to be favorable. 
For group V, VI, and VII elements, formation energies increase as we move down in the periodic table. 
This is consistent with the recent demonstration of plasma doping of MoS$_2$ with N~\cite{azcatl2016covalent}. 
The formation energies associated with doping the chalcogen site with group III and IV elements are rather constant. 
The overall strain induced to the lattice by chalcogen doping is slightly compressive, and constant over the entire range of substituents.
Finally, we found almost no difference between the energetics and strain of molybdenum and tungsten dichalcogenides.

We note that negative formation energies are not required for doping to be possible since experimental conditions during synthesis can change the chemical
potentials of the elements and, consequently, the formation energies.
Here we assume that the TMD is isolated and doping occurs between a free-standing sheet and the dopant element taken in its standard conditions.
Doping has been reported in various phases; liquid~\cite{yang2014chloride}, gas~\cite{suh2014doping} and recently plasma~\cite{azcatl2016covalent}, conditions 
that would strongly affect the chemical potential of the dopant.
Additionally, it has been demonstrated that the interaction between the TMD and its substrate also plays an important role to doping~\cite{zhang2015manganese}.
We study further the stability of some key compounds as a function of the chemical potential of its component in Section~\ref{sec:subvsint}.

\subsection{Electronic structure and potential substitutional dopants} 

The principal reason for doping semiconductors is to tune their electronic structure in order to control carrier conduction.
To identify potential n-type and p-type dopants, we analyzed the electronic density of state of all doped TMDs the automatic procedure described above.
Figure \ref{fig:fig3} shows the conduction band minima (green line) and Fermi energy (red) referenced to the top of the valence band (blue) 
across the entire set of elements tested. Vertical lines in the bandgap indicate defect states. 
The figures corresponding to MoSe$_2$, WS$_2$ and WSe$_2$ are reported in the Supplementary Material, Figures S4-S6.
Whether we consider sulfides or selenides, the overall trends including the variation of the bandgap, the Fermi energy and the number of defects are similar.
We note however that bandgaps reduce from sulfides to selenides.

\begin{figure}[!ht]
\centering
  \includegraphics[width=0.95\textwidth]{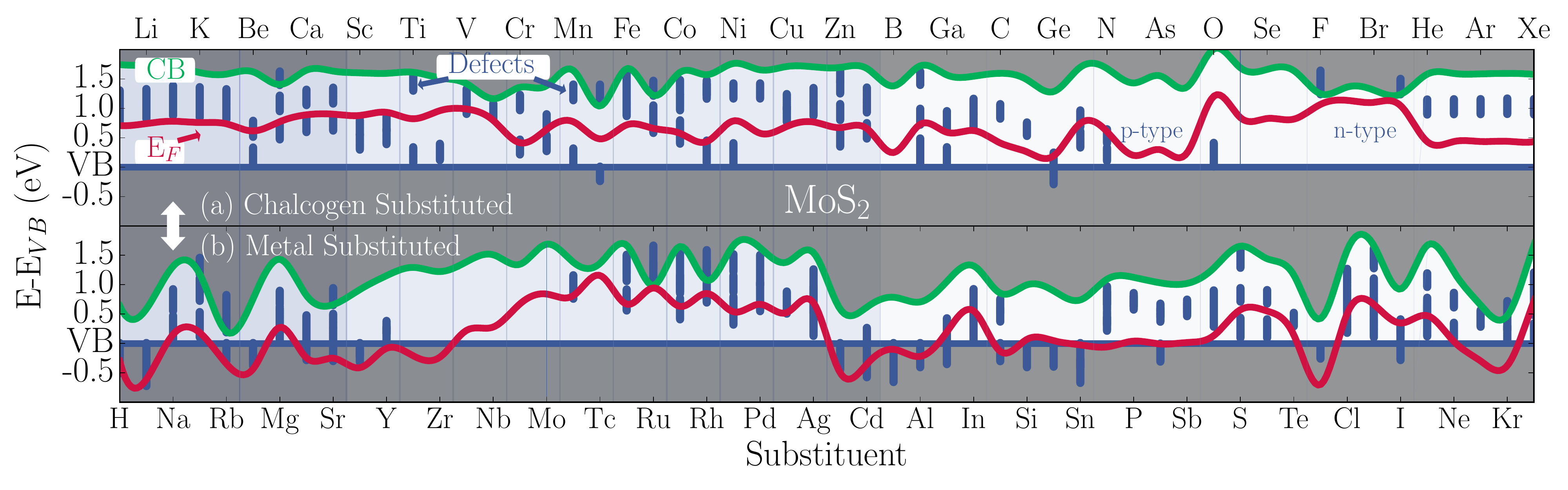}
  \caption{Energy of the conduction band (CB, green), fermi level (EF, red) and defects in the bandgap (BG, blue) with respect to the energy of the valence band 
  (VB, blue line x = 0) of S- (a: top) and Mo- (b: bottom) substituted monolayer MoS$_2$ as a function of substituents.}
  \label{fig:fig4}
\end{figure}

We start the discussion with the effect of substitutional doping of the chalcogen atom to the electronic structure (top panel of Figure \ref{fig:fig4}a).
We found that p-doping can be achieved by incorporation of group V elements in the X-sites, while halogens result in n-type. 
These results are consistent with prior simulations~\cite{yue2013functionalization,dolui2013possible} and we note that our automated analysis captures subtleties 
in the electronic DoS such as the presence of localized states at the Fermi energy right below the conduction band for halogens (see for example F and I). 
Moreover, our results are consistent with experimental data of N~\cite{azcatl2016covalent} and Cl~\cite{yang2014chloride} doping of MoS$_2$ resulting in p-doping and n-doping, respectively.
In fact, our simulations show that doping MoSe$_2$ or WSe$_2$ with N should result in improved performance due to a significantly reduced density of gap states.
However, we predict a 20\% reduction of the bandgap when go down the chalcogen column S$\rightarrow$Se. 
Importantly, doping with B is also predicted to lead to p-type doping for all four cases and one can tune the bandgap by choosing the nature of the TMD.
For example, we predict bandgaps of 1.42, 1.38, 1.31 and 1.17 eV for WS$_2$, MoS$_2$, WSe$_2$ and MoSe$_2$ chalcogen substituted with B, respectively.
Additionally, some group IV elements also result in p-behavior (Si, Ge, Sn).

Quite interestingly, substitution of the TM leads to significant flexibility both for p- and n-doping. 
We now focus on the bottom panels of the Figure \ref{fig:fig4}b.
The predicted bandgap across the TM family exhibits a maximum for the mid-TMs with reduction into the late and early TMs. 
This trend can be explained, to a large degree, by the induced strain as tension is known to reduce the bandgap. 
Thus, the atomic radius of the substituent governs the strain and, as a consequence, the trends in bandgap. 
The Fermi energy is lowered to the valence band by substitution with early TMs, leading to p-type behavior (Sc, Ti, Zr, V, Nb),  which has been
already demonstrated experimentally with Nb-doped MoS$_2$~\cite{suh2014doping}.
As the d-shell is filled, the Fermi energy increases relative to the conduction and valence band edges and we observe a transition from p to n-type doping for mid 
TMs (Tc, Ru, Rh). 
Doping with late TMs results in a large number of gap states and, thus, we expect poor transport properties.
For example, the substitution of Mo in MoS$_2$ with Mn creates defects at the Fermi energy, consistent with previous measurements~\cite{zhang2015manganese}.

\subsection{Accuracy of the GGA calculations} 

It is well known that the PBE exchange-correlation potential underestimates bandgaps and, various correction schemes have been proposed.
Reports show that the GW scheme overestimates bandgaps of monolayer TMDs~\cite{ramasubramaniam2012large} and DFT+U requires 
different values of U to be used depending on the nature of the dopant~\cite{andriotis2014tunable}. 
The hybrid functional HSE~\cite{heyd2003hybrid}, including part of the exact Hartree-Fock exchange, provides the best agreement with experiments, 
especially to describe the atomic and electronic structure of bulk~\cite{peelaers2014first} and some monolayer TMDs~\cite{hu2015electronic}. 
Our calculations of the HSE bandgaps of free-standing MoS$_2$, MoSe$_2$ and MoTe$_2$ monolayers are 2.00, 1.91 and 1.48 eV, respectively. 
These values are larger than reported measurements of 1.90, 1.58 and 1.10 eV~\cite{mak2010atomically,zhang2014direct,ruppert2014optical}.
On the other hand, the PBE functional tends to underestimate bandgaps (see Table S3 for details).
We selected a subgroup of the substituents presented above for HSE calculations to evaluate model uncertainties.
We fully relaxed various doped TMDs with the HSE functional including pristine MoS$_2$ and, chalcogen and TM substituted MoS$_2$ with Cl (as well as N) 
and Nb, respectively. 
We found a very good agreement between atomic structures computed with PBE and HSE functionals.
Interestingly, formation energies computed with PBE are similar to that computed with hybrid functional with less than 0.3 eV error. 

\begin{figure}[!ht]
  \centering
  \includegraphics[width=0.45\textwidth]{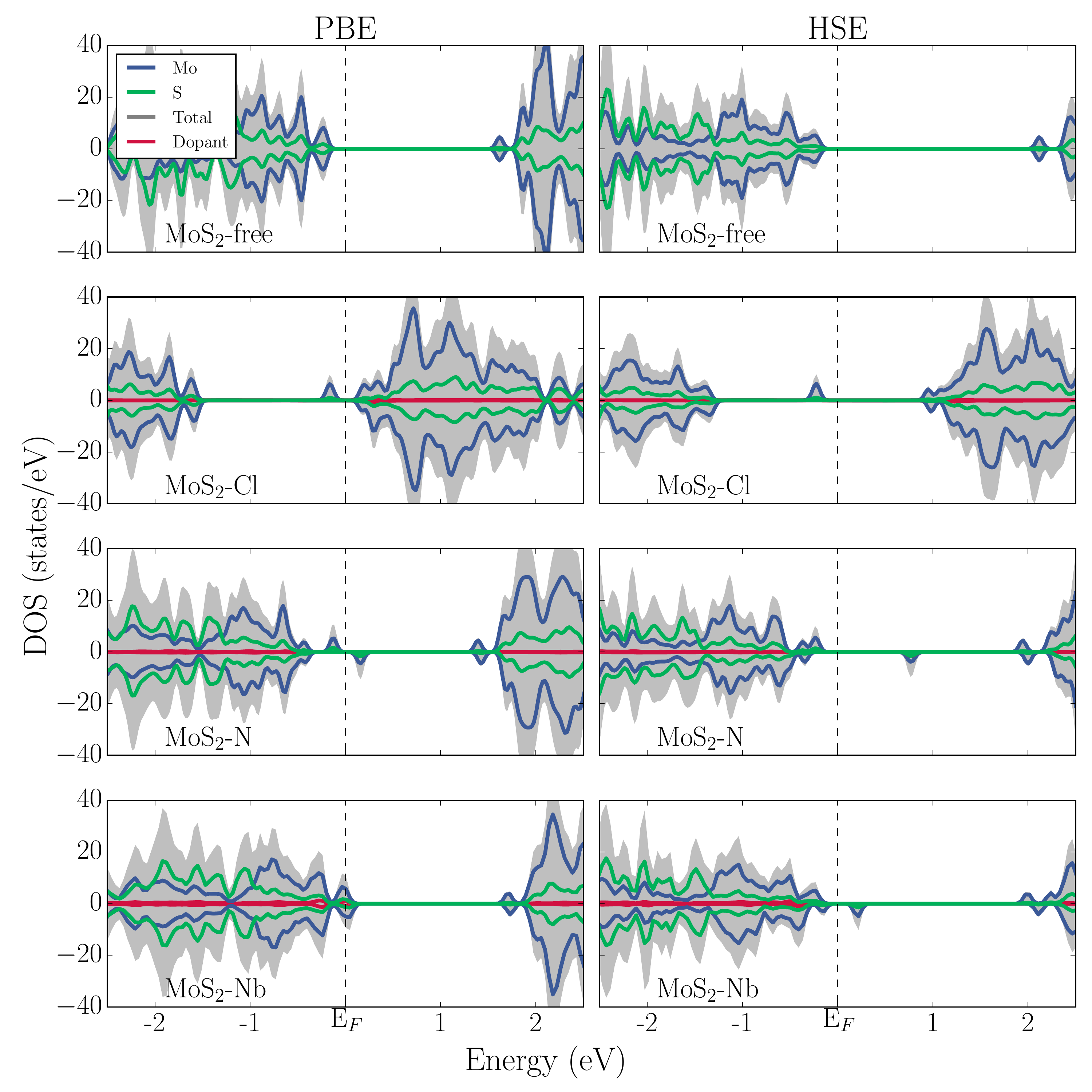}
  \caption{Projected density of state of pristine MoS$_2$ as well as various chalcogen and TM substituted MoS$_2$ computed with PBE (left) and HSE (right) functionals.}
  \label{fig:fig5}
\end{figure}

Figure \ref{fig:fig5} shows the electronic DoS computed with PBE and HSE.
Overall we find that the main features and trends observed in GGA are confirmed by the HSE calculations.
For pristine MoX$_2$ (with X = S, Se), PBE bandgaps are approximatively 20\% smaller than experiments whereas HSE bandgaps are 15\% larger, 
consistent with recent reports~\cite{ozcelik2016band}.
Defect states calculated with HSE also appear slightly shifted in energy compared to those calculated with PBE; we note larger energy splitting between 
defect states in different spin states. This has been observed before and our DoS agree with previous work~\cite{hu2015electronic}. 
To summarize, we acknowledge that the band plots presented Figure \ref{fig:fig4} (as well as S4-S6) underestimate bandgaps however, we are confident 
in the prediction of the number of defect states and Fermi energy.
Such a high-throughput study has been possible only because of the reduced computational cost and fast convergency of PBE calculations.

\section{Interstitial sites}
\label{sec:int}

\subsection{Energetics of absorbed versus interstitial sites}

\begin{figure}[!ht]
\centering
(a)\includegraphics[width=0.95\textwidth]{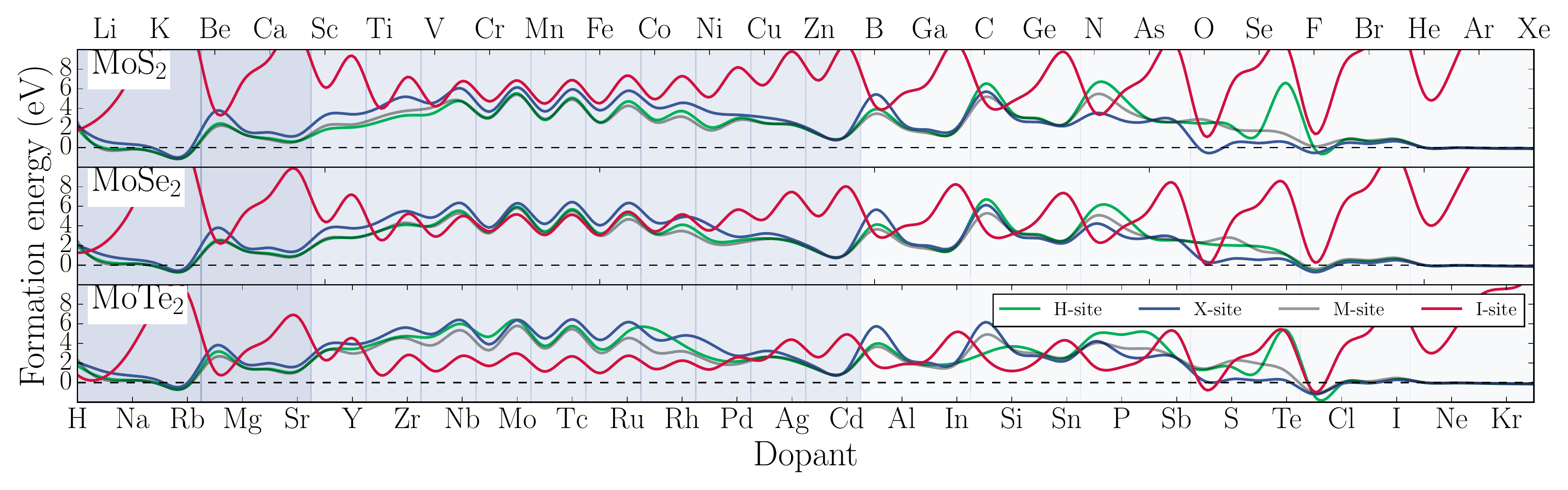}
(b)\includegraphics[width=0.95\textwidth]{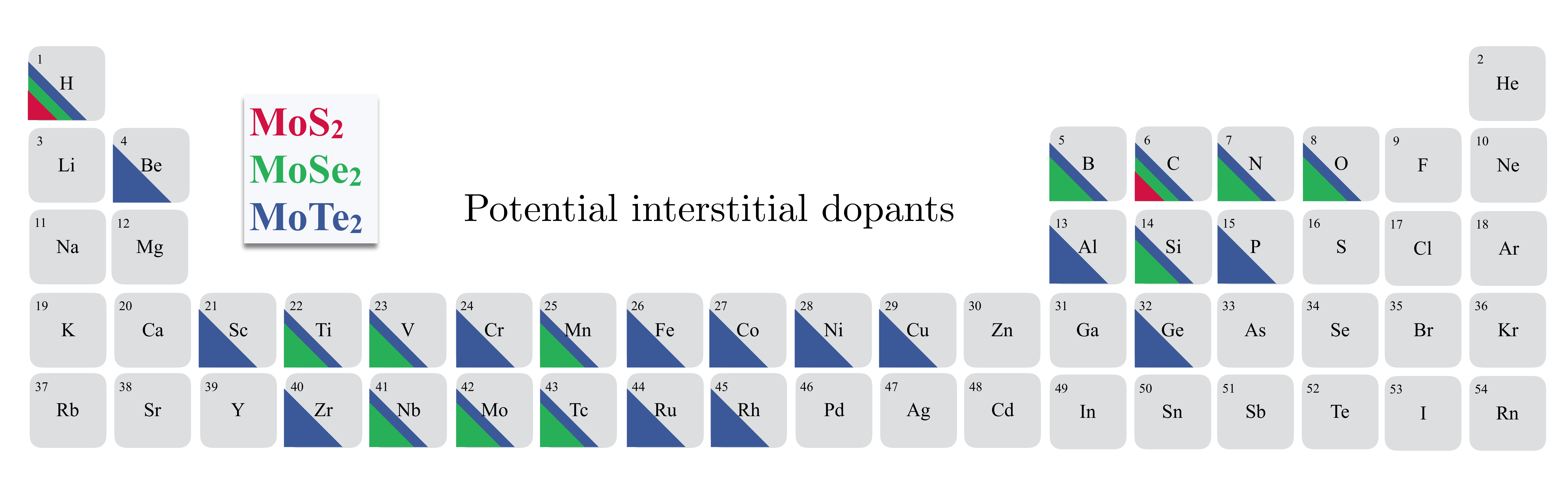}
\caption{Formation energy of dopants adsorbed at various sites (H, M and X) on the surface of molybdenum dichalcogenides as well as at interstitial (I) sites (a).
Potential interstitial dopants in MoS$_2$, MoSe$_2$ and MoTe$_2$ predicted from our analysis (b).}
\label{fig:fig6}
\end{figure}

Figure \ref{fig:fig6}a shows formation energies for the entire set of dopants studied in the three adsorption sites and the I interstitial site for molybdenum disulfide, diselenide, and
ditelluride.
Quite surprisingly a significant fraction of the dopants studied favor the interstitial site over the adsorbed ones. Few elements favor the I' site over I (for example Cu in MoTe$_2$) and 
the energy difference between the two is small. Therefore, in the following we will only consider interstitial sites located at the symmetry plane of the TMD (I-site) for simplicity.
Figure S9 contains the complete energetic analysis.
The adsorption energies (H, X and M curves) are rather independent of the nature of the chalcogen however, we find a strong dependence of the interstitial energies (I curves)
on the nature and lattice parameter of the TMD.
The number of elements that energetically prefer interstitial sites over adsorbed increases from sulfides to selenides and to tellurides, as summarized in the periodic
table of Figure \ref{fig:fig2}b.
Only two elements (H and C) have a ground state at interstitial site in the well-studied MoS$_2$. Interestingly, twelve elements in MoSe$_2$ and more than half the elements tested 
in MoTe$_2$ are predicted to energetically prefer the interstitial sites. Potential interstitial dopants in MoTe$_2$ include most of the transition metals (TMs) as well as some first and 
second row (2p/3p) elements and, contain  that of MoSe$_2$ which contain that of MoS$_2$.

\begin{figure}[!ht]
\centering
\includegraphics[width=0.95\textwidth]{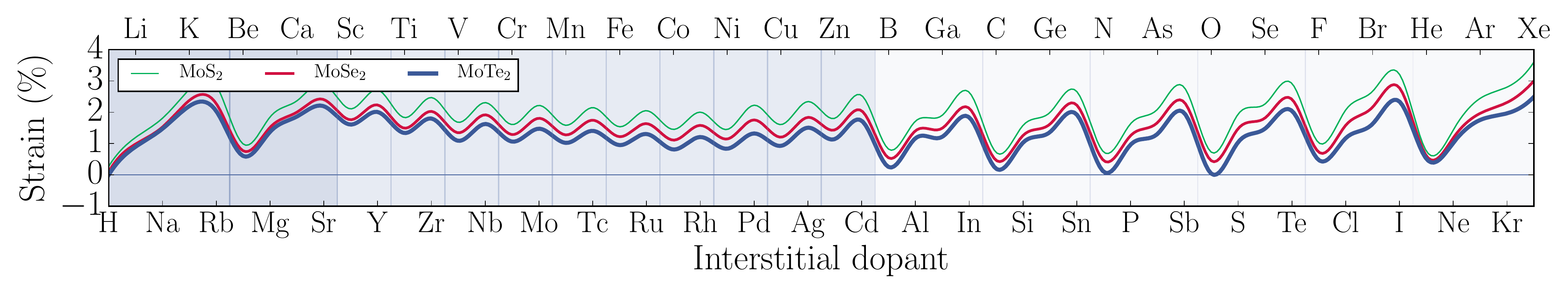}
\caption{Biaxial strain induced to the lattice of MoS$_2$, MoSe$_2$ and MoTe$_2$ upon interstitial doping.}
\label{fig:fig7}
\end{figure}

Figure \ref{fig:fig7} shows the strain induced to the lattice of MoS$_2$, MoSe$_2$ and MoTe$_2$ upon interstitial doping.
The strain decreases from sulfides to tellurides, consistent with the increasing size of the lattice parameter of the corresponding host TMD.
Moreover, strain correlates with the size of the dopant and the trends in formation energy of interstitial doping; increasing as we move down the periodic table 
(i.e. increasing atomic radius of the dopant). 
Interestingly, light elements (H, B, C, N, O, F and He) barely strain the lattice of MoTe$_2$ ($<0.2$ \%) suggesting the possibility of high concentration 
interstitial doping.

\begin{figure}[!ht]
\centering
\includegraphics[width=0.45\textwidth]{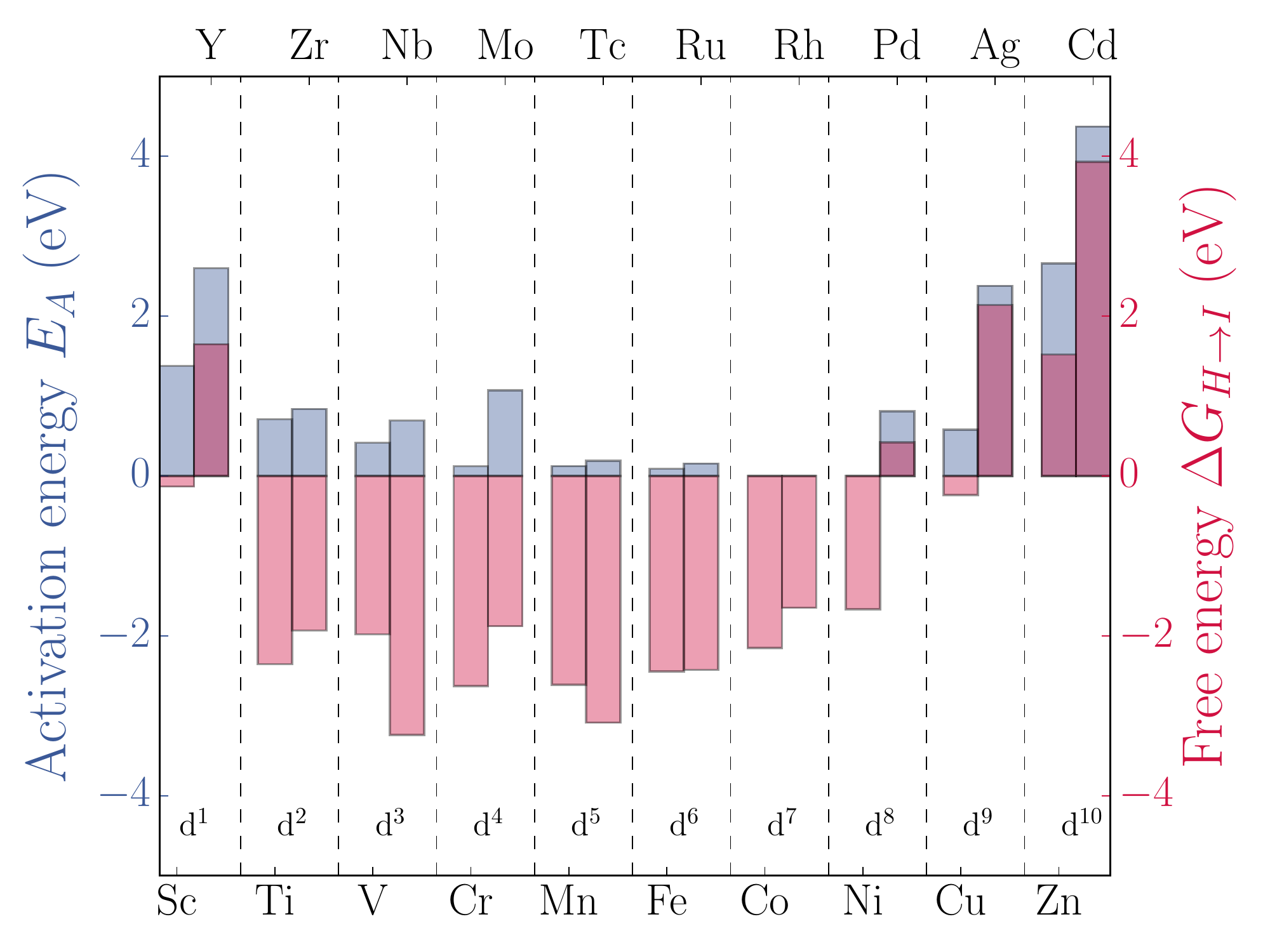}
\caption{Activation energy and free energy corresponding to interstitial doping MoTe$_2$ with TMs.}
\label{fig:fig8}
\end{figure}

Figure~\ref{fig:fig6} can be used to asses the thermodynamics of interstitial doping, however it lacks information about the kinetics of the doping reaction.
The potential energy surface upon interstitial doping presented Figure~\ref{fig:fig1} provides both kinetics and thermodynamics information.
Since TMs are promising for interstitial doping, we performed NEB calculations of all TMs to obtain the minimum energy path associated with 
moving the dopant from one surface of the  MoTe$_2$ monolayer to the opposite side. These curves provide information about the activation energy 
E$_A$ and free energy for interstitial doping $\Delta G_{H\rightarrow I} = E_I-E_H$. These results are summarized Figure \ref{fig:fig8} and reported Table S4.
Only relatively small activation barriers will be accessible via standard doping techniques and, the kinetics of the doping reaction will depend on them. 
Beside $d^1$ and $d^{10}$ TMs, the energy barriers for doping are small and the corresponding free energies are mostly negative, suggesting
favorable doping reaction.
Both activation barrier and free energy for doping describe an approximate parabolic shape along the d-shell filling with a minimum around mid-shell similar to the variation 
of TMs' atomic radius.
Moreover, $3d$ elements exhibit lower activation barriers than their $4d$, more voluminous counterpart.
Interestingly, we predict negligible energy barriers ($<$ 0.2 eV) when doping occurs with elements Cr, Mn$\rightarrow$Ru and even no barrier for
Co$\rightarrow$Ni.
From the analysis of the entire NEB paths, we note the existence of I' sites for late TMs ranging from Ni$\rightarrow$Ag which represent the ground 
state among all doping sites in the case of Ni and Cu dopants. For more details, we provide the paths of TMs crossing free-standing MoTe$_2$ Figure S10.

Finally, we performed AIMD simulations of Nb and Cu interstitial in MoTe$_2$ at 300 and 500 K to further investigate  the stability of these
two representative compounds. 
The total energy as well as some snapshots of the atomic structure are reported Figure S11.
Overall, we find that both doped structures conserve their general geometry along the 15 ps of simulations at temperatures up to 500 K, 
indicating good thermal stability.
According to the energy landscape for doping Cu and Nb interstitial in MoTe$_2$  we predict barriers for the interstitial atom to escape the 
structure ($E_A-\Delta G_{H\rightarrow I}$) equal to 0.82 and 3.93 eV, respectively.
We note that Cu interstitial rapidly hops from its initial I-site to an I'-site at both 300 and 500 K, consistent with the predicted potential energy 
surface presented Figure \ref{fig:fig1}.
Interestingly, a Mo atom located next to the Nb interstitial hops from its lattice position to a neighboring interstitial site at 500 K. 
This mechanism releases some of the local stress generated by the interstitial Nb atom resulting in a Mo-vacancy between two interstitial atoms
(bottom left snapshot).

\subsection{Electronic structures and potential interstitial dopants}

\begin{figure}[!ht]
\centering
\includegraphics[width=0.95\textwidth]{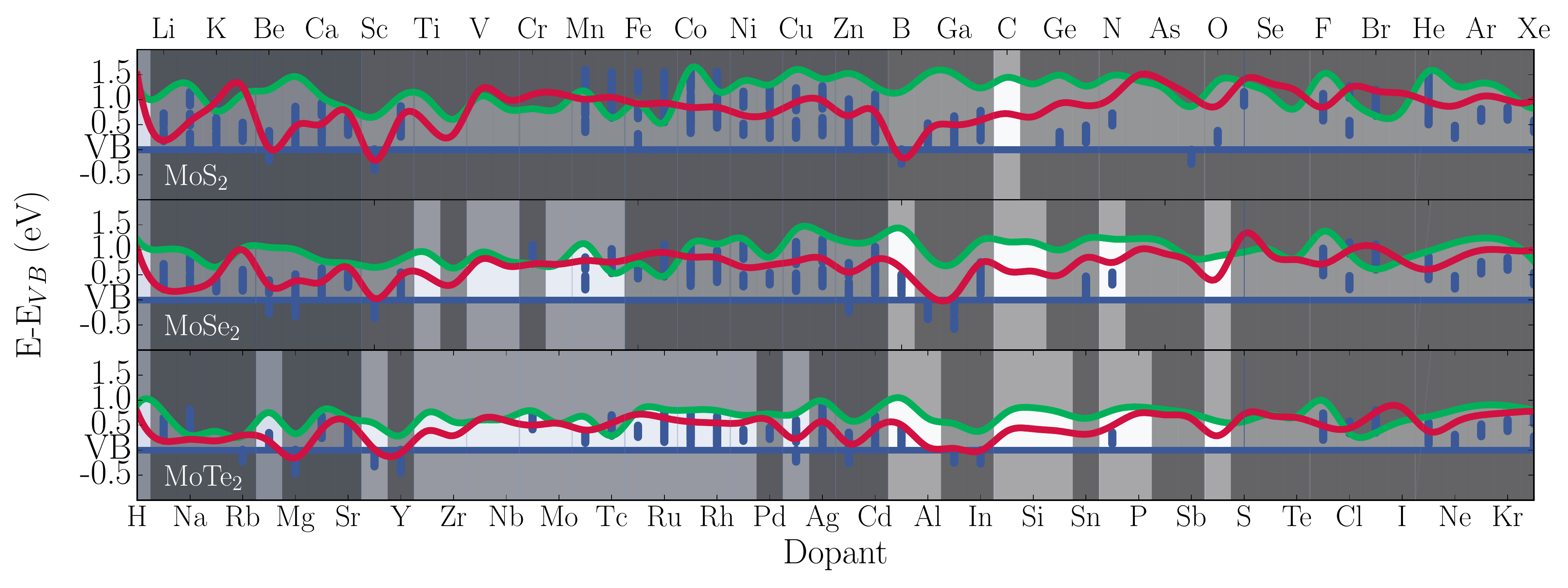}
\caption{Energy of the conduction band (CB, green), fermi level (E$_F$, red) and defects in the band gap (BG, blue) with respect to the energy of the valence 
band (VB, blue line x = 0) of MoTe$_2$ doped at M, H, X and I sites as a function of dopant. 
Light regions correspond to configuration with the lowest dopant formation energy.}
\label{fig:fig9}
\end{figure}

Figure~\ref{fig:fig9} shows the (PBE) conduction band energy (green), Fermi energy (red) and gap states (vertical blue lines) with respect to the 
valence band (VB) energy (blue horizontal line) for interstitial doping of MoS$_2$, MoSe$_2$, and MoTe$_2$ with all the element tested. 
Light regions highlight dopants for which the interstitial site is favored energetically as compared to
surface states, as discussed in the previous section. The corresponding figures for the adsorbed sites are reported Figures S12-S14. 
The overall trends including bandgap, Fermi energy and the number of defect states among the three dichalcogenides studied are 
similar. We note that bandgaps reduce from disulfides to ditellurides and, as explained above, the number of interstitial dopants increase with 
increasing lattice parameter of the TMD.

We will focus on MoTe$_2$ as a representative compound because of the large number of predicted low energy interstitial dopants.
It is known that various interstitial or adsorbed dopants can shift the Fermi energy to the valence band maximum (corresponding to p-type conduction)
or to the conduction band (n-type), depending on their acceptor/donor character. 
However, this simple rule is only qualitative and the Fermi level shift rigorously depends on the complex hybridization between the orbitals of the 
dopant and that of the TMD host. 
According to Figure~\ref{fig:fig9}, only Sc and Al dopants induce p-type conduction.
We note a defect state close to the VB in the case of Sc.
Early to mid-TMs (V$\rightarrow$Mo) show n-type conduction with few or no defect states.
Late TMs present numerous defects in the bandgap precluding efficient electronic application.
Beside TMs, only phosphorus appears as a good n-type interstitial dopant.
Our analysis also provides information related to the Fermi level shifting of various adsorbed elements. 
For instance, the adsorption of alkali/alkali-earth metals leads to n-type conduction whereas adsorbed halogens leads to p-type conduction
(consistent with Ref.~\cite{dolui2013possible}).

\begin{figure}[!ht]
\centering                                                                                                                                                                                                                                                                                                                                                                                                                                                                                                                                                                                                                                                                                                                                                                                                                                                                                                                                                                                                                                                                                                                                                                                                                                                                                                                                              
\includegraphics[width=0.45\textwidth]{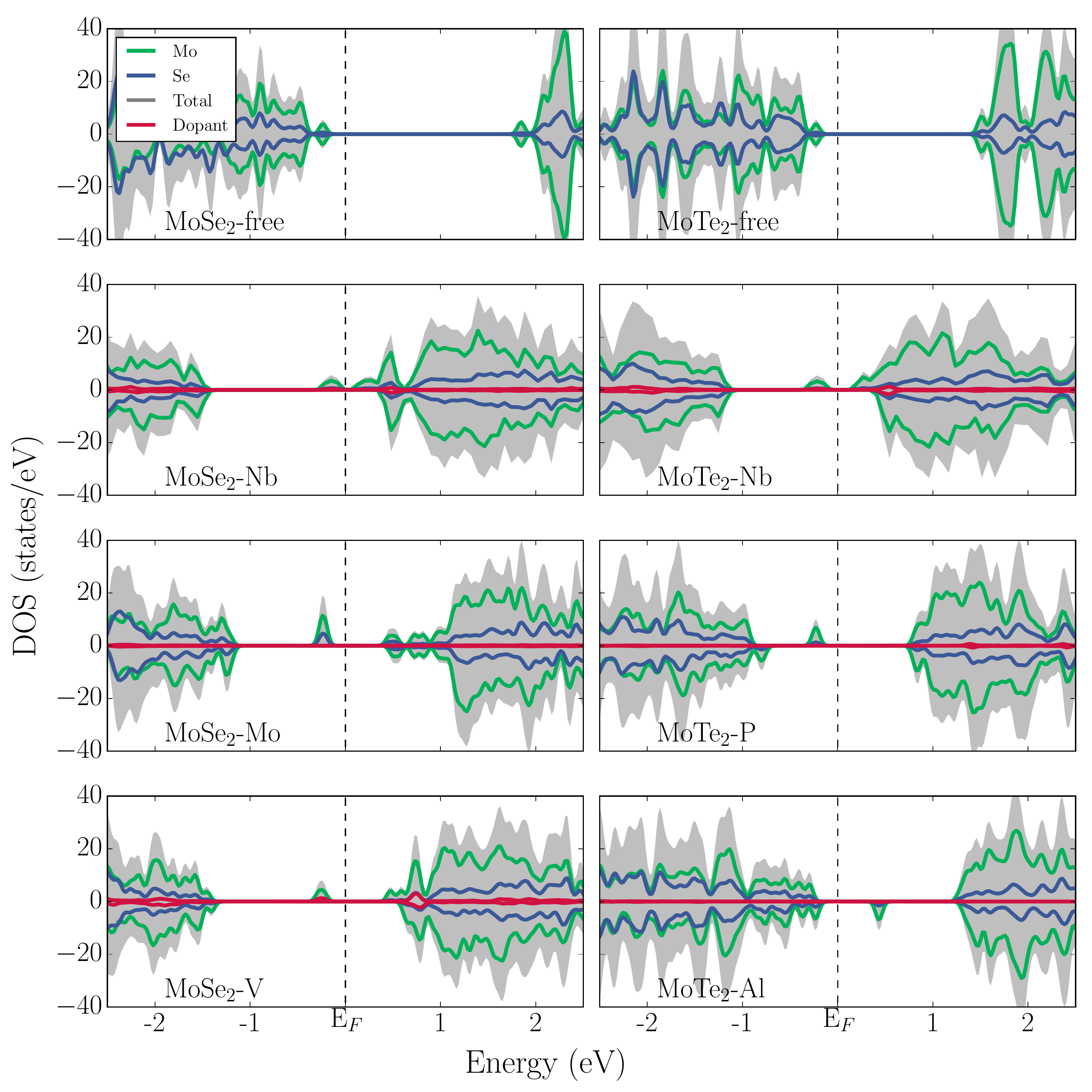}
\caption{Total (grey) and projected density of states over metal (green), chalcogen (blue) and interstitial dopant (red) of 
molybdenum diselenides (left) and ditellurides (right) with and without an impurity atom at interstitial site.}
\label{fig:fig10}
\end{figure}

We now discuss with additional details the electronic structure corresponding to some of the most interesting doped cases. 
Figure~\ref{fig:fig10} shows the projected density of states of some key interstitial-doped MoSe$_2$ and MoTe$_2$ computed at the HSE level. 
Based on the discussion above, these HSE results should be considered as an upper limit for bandgap size; all PDoS computed with PBE, 
representing a lower limit, can be found for comparison on our \href{https://nanohub.org/labs/run/pdos/}{website}.

All doped TMDs reported in Figure~\ref{fig:fig10} exhibit a magnetic ground states with a magnetic moment of 1$\mu_B$ except in the case of Mo-doping with 
2$\mu_B$. Interstitial doping MoSe$_2$ and MoTe$_2$ with Nb (which is energetically favorable) shifts the Fermi energy towards the conduction band leading 
to n-type conduction. We note the existence of a Nb occupied defect state located approximately 0.3 eV below the conduction band. 
Similar n-type conduction can be achieved by doping these TMDs with Mo and V however we observe larger energy splitting of the dopant defect state
moving approximately 0.6 eV into the band gap.
An occupied defect state is observed deeper into the band gap when MoTe$_2$ is doped with P, located approximately 1.0 eV below the conduction band.
Although, an interstitial Mo atom in MoTe$_2$ has a relatively low formation energy of 3 eV, it has been shown that the probability of forming such defect 
during chemical growth conditions is low~\cite{haldar2015systematic}.
Mo-interstitials are predicted to shift the Fermi energy toward the conduction band and contribute to the observed n-type character of these TMDs 
mainly attributed to chalcogen and dichalcogen vacancies.
Finally, interstitial doping of MoTe$_2$ with Al costs 2.0 eV with respect to the fcc metallic phase, and leads to p-type conduction with 
an unoccupied defect state located approximately 0.6 eV above the valence band.
	
\subsection{Simulated STM images of interstitial doping in MoTe$_2$}

\begin{figure}[!ht]
\centering
\includegraphics[width=0.95\textwidth]{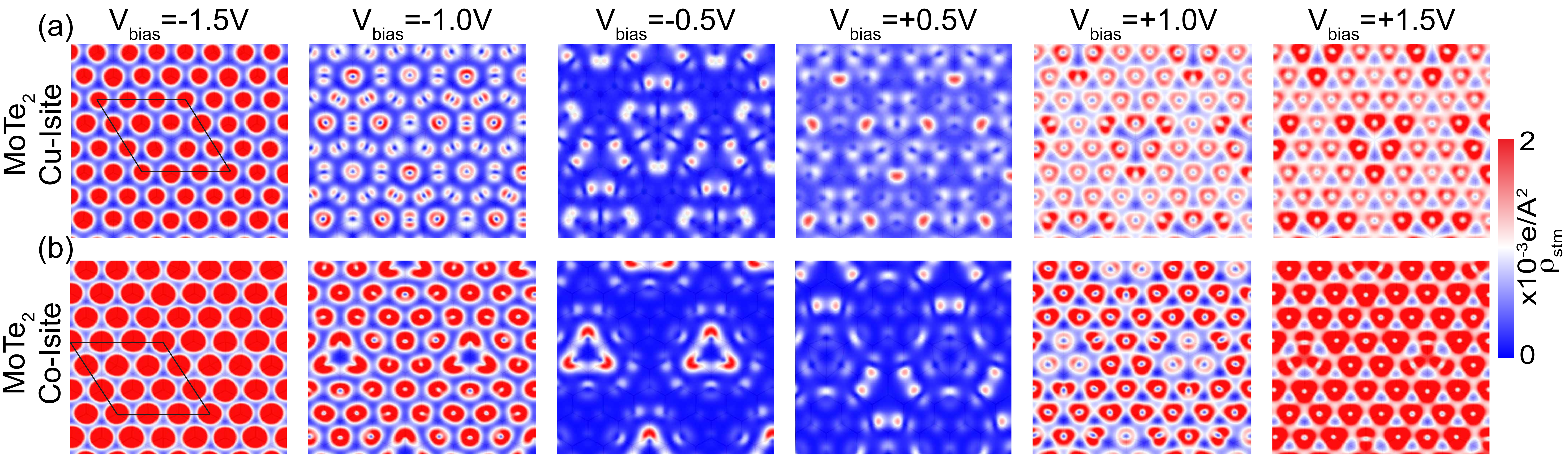}
\caption{Simulated STM images of interstitial Cu (a) and Co (b) in MoTe$_2$. 
The $\rho_{STM}$ is shown at constant height, corresponding to STM current imaging mode for different values of voltage bias ($V_{bias}$). 
The supercell is shown in with black lines.}
\label{fig:fig11}
\end{figure}

\begin{figure}[!ht]
\centering
\includegraphics[width=0.95\textwidth]{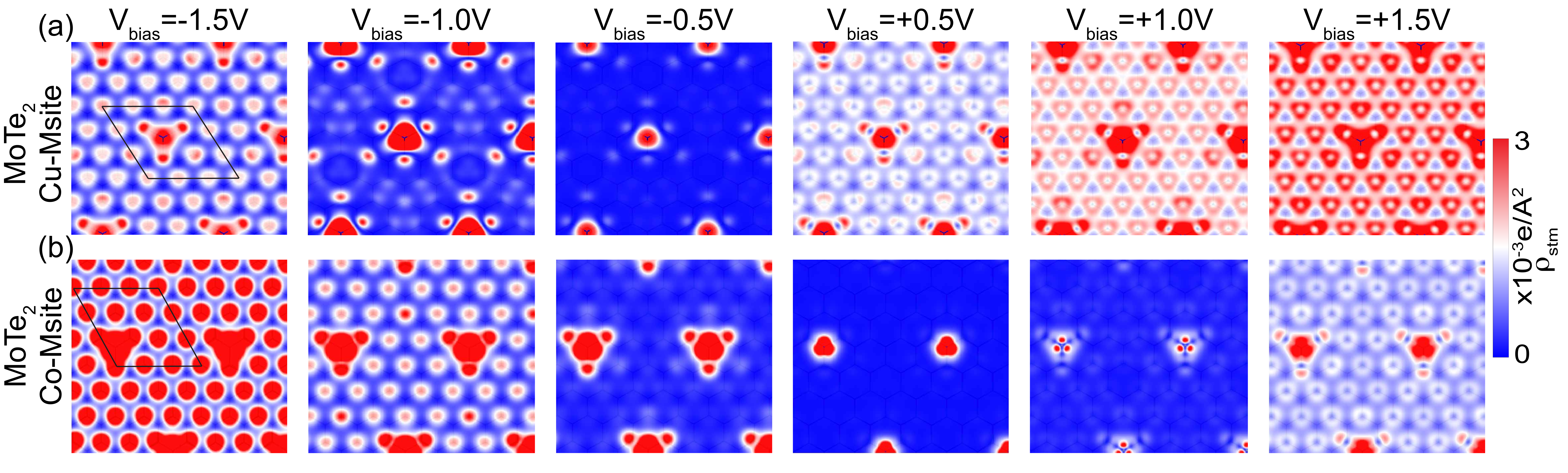}
\caption{Simulated STM images of Cu (a) and Co (b) adatoms at M-sites in MoTe$_2$. 
The $\rho_{STM}$ is shown at constant height, corresponding to STM current imaging mode for different values of voltage bias ($V_{bias}$). 
The supercell is shown in with black lines.}
\label{fig:fig12}
\end{figure}

Given that interstitial doping has not been reported experimentally, we use the atomic structures computed from first principles calculations to predict
scanning tunnelling microscopy (STM) images of some key compounds.
STM imaging is a powerful technique to identify defects and structural transformations in TMDs~\cite{Zhang2017,Chiu2015,Park2016}. 
Yet, assignment of atomistic structure to STM images can be challenging, simulated STM images can serve as guide to experimentalist to aid the identification 
of the doping sites and species, such as the one proposed in this work. 
Simulated STM images of copper and cobalt dopants at M and I sites in MoTe$_2$ are shown Figures \ref{fig:fig11} and \ref{fig:fig12}. 
Details on the calculation are presented Section S4.
The contour plots represent the charge density $\rho_{STM}(r, V_{bias})$~\cite{tersoff1983theory,tersoff1985theory}, as computed by Eq. S1, 
for various bias voltages ranging between -1.5 V and +1.5 V. 
The STM images correspond to a tip-to-sample distance of $d=2.5$ \AA~measured from the topmost atom of the doped MoTe$_2$ monolayer.

At large reverse bias the STM image of MoTe$_2$ doped at the I-site shows similar patterns as the pristine monolayer, that is, high density regions with circular shape at 
the topmost (Te) atoms and, low density regions at the transition metal sites (see Figure \ref{fig:fig11}). 
Therefore, dopants at I-site, lying underneath the topmost chalcogen plane, cannot be identified at large reverse bias voltages. 
By contrast, interstitial Cu in MoTe$_2$ can clearly be identified at forward bias voltages for which the STM images show high density regions with triangular shape 
around the locally distorted structure (due to the presence of copper), see Figure \ref{fig:fig11}a for $V_{bias}=$+1 and +1.5 V.      
Tellurium lattice distortion induced by the cobalt at the I-site can be identified at $V_{bias}$=-1 V for which the STM image shows 
high density regions around bonds (bonding-like features) and hexagonal-shaped low density regions at the cobalt site, see Figure \ref{fig:fig11}b.

In the case of dopants occupying M-sites (Figure \ref{fig:fig12}) we observed that the highest values of $\rho_{STM}$ correspond to the location of adatoms which 
are the topmost species on the surface. 
The underlying hexagonal structure of the TMD is better resolved at high bias voltages in Cu-doped MoTe$_2$ at M-sites than that for Co-doping. 
This is explained by the fact that Cu is adsorbed at a longer distance from the first plane of Te atoms than Co ($d_{\text{Cu-Te}}=1.34$ \AA~and 
$d_{\text{Co-Te}}=0.86$ \AA), hence requiring stronger bias voltage for the tunnelling current to resolve the MoTe$_2$ matrix. 
Due to the fact that Cu and Co at M-sites are the topmost atoms, low bias voltages are enough to identify the adatoms, as shown in Figure~\ref{fig:fig12}a and~\ref{fig:fig12}b 
for $V_{bias}=\pm$0.5 V.
To summarize, we found that simulated STM images of interstitial atoms present specific patterns, different from adsorbed sites which can be used to guide 
the identification of dopants in MoTe$_2$. 

\section{Discussion}

\subsection{Substitutional versus interstitial doping}
\label{sec:subvsint}

We now discuss the relative stability of the various potential doping sites as a function of the chemical potentials of the metal, chalcogen and dopant.
This can be achieved by a careful inspection of Figures~\ref{fig:fig2} and~\ref{fig:fig6} however, the large amount of information makes the task difficult.
Moreover, we remind the reader that substitutional doping was computed at the PBE level whereas interstitial calculations included Grimme's D2 correction.
Even though we do not expect a large energy difference due to the vdW correction, we relaxed some of the most interesting interstitial dopants at the PBE level for better comparison.
Our goal is to identify energetically favorable configuration as a function of the chemical potential of the metal, chalcogen and dopant.
In order to reduce the number of independent variables to two, we linearly vary the chemical potential of the chalcogen and metal atom between
the metal rich and chalcogen rich values. 
Thus, we define both chemical potentials as a function of a variable, $\xi$:

\begin{equation}
\label{eq:eq4}
  \begin{cases}
\mu_X(\xi) = \mu_X^0 \xi + \mu_X^{M-rich} (1-\xi) \\
\mu_M(\xi) = \mu_M^0 (1-\xi) + \mu_M^{X-rich} \xi
  \end{cases}
\end{equation}

\begin{figure}[!ht]
\centering
\includegraphics[width=0.65\textwidth]{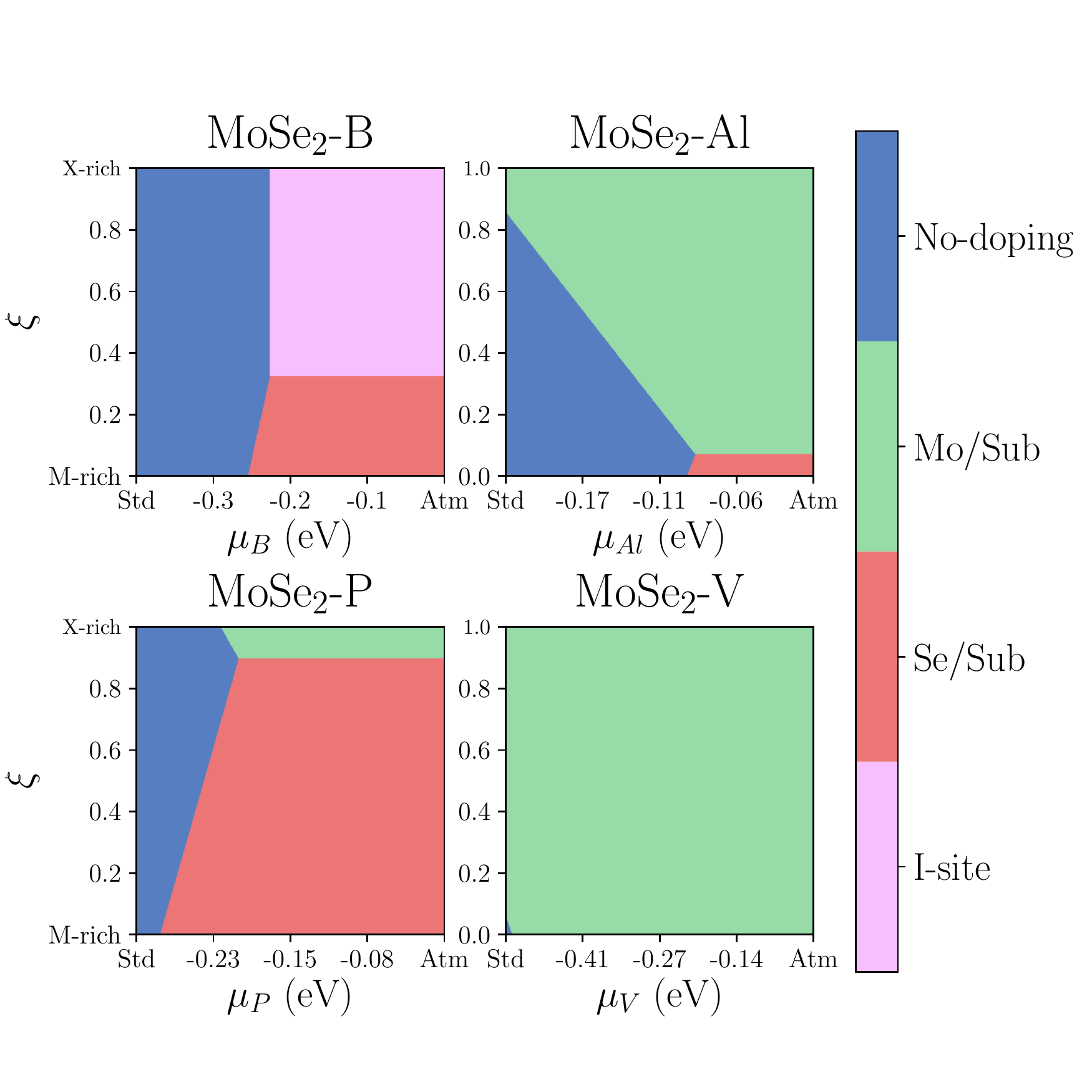}
\caption{Phase diagram of the various doping configuration studied for B, Al, P and V doped MoSe$_2$ as a function of the chemical potential of the metal, chalcogen
and dopant. The chemical potential of the dopant was taken between its standard (std) and atomic (atm) conditions.}
\label{fig:fig13}
\end{figure}

Eqs.~\ref{eq:eq4} leads to X-rich conditions when $\xi = 1$ and M-rich when $\xi = 0$.
The second variable is the chemical potential of the dopant ($\mu_D$) which varies between its standard and atomic conditions.
Figure~\ref{fig:fig13} shows the stability of the various configurations as a function of $\xi$ and $\mu_D$ in MoSe$_2$.
Regions of the chemical potential phase diagram are colored according to the lowest formation energy among the configurations:
Se substituted, Mo substituted and interstitial (I-site). Positive formation energies results in no-doping of the TMD.

We found that V almost always favors the substitution of the TM site regardless of its chemical potential and that of Mo and Se.
Only a small region of the chemical potential phase diagram results in no doping when the chemical potential of the dopant is close to its 
standard (Vanadium BCC) condition in the metal rich limit.
Similarly, doping MoSe$_2$ with Al results in Mo substitution however, we note the possibility of Se substitution when the chemical potential of Al
increases toward its atomic limit, in metal rich conditions.
Doping with phosphorous favors the substitution of the chalcogen site with the possibility of doping the TM site in Se rich conditions.
We found that B dopant favors the interstitial site for a wide range of chemical potentials.
The substitution of Se in MoSe$_2$ by B can be achieved in metal rich conditions. 
Although few dopants favor the interstitial site in MoSe$_2$ (we predict H, B and C), we expect many more in MoTe$_2$.

\subsection{Role of possible compensating defects} 

Compensating defects can have a strong effect on the ability of engineered defects to change electrical characteristics~\cite{yang2015self,alberi2016suppression}.
In our case, the most likely culprits are TM and chalcogen vacancies and anti-site defects (TM on the chalcogen site and vice versa). 
We will base the discussion on MoS$_2$ as a representative material, we expect similar behavior in the other cases. 
The formation energy of Mo vacancies is high (3.7 eV in S-rich conditions and 7.4 eV in Mo-rich conditions) and, thus, is likely unimportant. 
A sulfur vacancy, V$_S$, in the neutral state results in an empty state about 0.5 eV below the conduction band minima that can trap electrons; 
thus, limiting n-type doping. Their formation energies (3.2 eV in S-rich conditions and 1.4 eV in Mo-rich conditions) indicate they will be more 
preponderant that V$_{Mo}$ but healing the vacancy with a halogen would reduce the energy and result in n-type doping. 
Anti-site defects are also relatively high in formation energies. 
The lowest being S in the Mo site under S-rich conditions with a formation energy of 2.6 eV and could be avoided with less S favorable conditions.

\section{Conclusion}

We showed that potential candidates for tuning the carrier density in TMDs are not limited to the substitution with elements 
that belong to neighboring electronic shell than that of the constituents. Our DFT simulations indicate that group III and IV elements should be explored 
experimentally to substitute chalcogen atoms and various early TMs are promising to replace Mo/W leading to p-type conduction.
Interestingly, we predict that a wide range of bandgaps can be achieved by selecting the nature of boron-doped TMDs. 
Moreover, we predict improved electronic properties of N-doped transition metal diselenides over sulfides owing to a reduction in gap states.
Furthermore, we found that the trends in formation energy and bandgaps of Mo-substituted TMDs correlate with the evolution of atomic radius 
of TM-dopants. 
Although single-layer TMDs are less than 1 nm thick, we demonstrated the possibility of impurity interstitial with a large fraction of the periodic table.
We identified 2, 12 and 25 elements (out of the 54 tested) having low energy at interstitial site in MoS$_2$, MoSe$_2$ and MoTe$_2$, respectively.
We explored potential interstitial dopants based on the following criteria: (i) lowest formation energy at interstitial site compared to elements adsorbed 
on the surface of the TMD;  (ii) small activation energy for doping and; (iii) the Fermi energy shift of the corresponding electronic structure.
The automatic analysis of the electronic structure summarized Figures~\ref{fig:fig9} showcases the band diagram of approximately 5,000
doped TMDs and, suggests promising candidates among most of the TMs in MoSe$_2$ and MoTe$_2$. 
For instance, Nb interstitial is predicted to achieve n-type conduction and we demonstrated a good stability of the compound.

We believe that the lack of experimental proof of the existence of interstitial doping is mainly due to the unfavorable energies of interstitial in the widely
studied MoS$_2$ combined with detection challenges.
Simulated STM images provide a footprint of interstitial dopant in MoTe$_2$ and we believe that our theoretical demonstration will motivate microscope experts 
to identify such impurity in TMDs and other 2D materials.
Additional properties should be explored to fully take advantage of this emerging doping technique in TMDs.
Finally, this study can be easily extended to other TMDs, phases and dopants.
For example, Er-doped MoS$_2$ has recently been achieved experimentally giving rise to near-infrared photoluminescence~\cite{bai20162d}.

\section*{Supplementary Material}
\label{sec:supp}
Sections S1 and S2 show the atomic and electronic structures of substituted and interstitially doped TMDs, respectively.
Section S3 and S4 include details on the effect of spin-orbit coupling to the doping energies and, the procedure to compute the simulated STM images.
The DoS of all doped TMDs are provided on our \href{https://nanohub.org/labs/run/pdos/}{website}.

\subsection*{Acknowledgement}
This work was partially supported by the FAME and LEAST Centers, two of six centers of STARnet, a Semiconductor Research Corporation program 
sponsored by MARCO and DARPA. We thank nanoHUB.org and Purdue for the computational resources.
 
\clearpage

\end{document}